\documentclass[%
mathleft,%
]{an}
\usepackage{graphicx}
\usepackage{times}

\newcommand{\NH}{$N_{\mathrm{H}}\,$}
\newcommand{\NHI}{$N_{\mathrm{HI}}\,$}
\newcommand{\NHmath}{$N_{\mathrm{H}}$}
\newcommand{\NHImath}{$N_{\mathrm{HI}}$}
\newcommand{\HI}{H{\sc I}\,}

\newcommand{\percmq}{cm$^{-2}\,$}

\newcommand{\nms}{\mathsurround=0pt}
\newcommand{\oversim}[2]{\lower 2pt\vbox{\baselineskip 0pt \lineskip 1pt \ialign{$\nms#1\hfil##\hfil$\crcr#2\crcr\sim\crcr}}}
\newcommand{\gtsim}{\mathrel{\mathpalette\oversim{>}}}
\newcommand{\ltsim}{\mathrel{\mathpalette\oversim{<}}}

\newcommand{\aap}{A\&A }
\newcommand{\aaps}{A\&AS }
\newcommand{\sci}{Science }

\begin{document}

\Pagespan{1}{}
\Yearpublication{2011}%
\Yearsubmission{2011}%
\Month{1}%
\Volume{999}%
\Issue{92}%

\title{\NH - \NHI correlation in Gigahertz-peaked-spectrum galaxies \,\thanks{Data from the Westerbork Synthesis Radio Telescope}}

\author{L. Ostorero\inst{1,2}\fnmsep\thanks{Corresponding author:\email{ostorero@ph.unito.it}}
\and  R. Morganti\inst{3,4}
\and  A. Diaferio\inst{1,2}
\and  A. Siemiginowska\inst{5}
\and  \L. Stawarz\inst{6}
\and  R. Moderski\inst{7}
\and  A. Labiano\inst{8}
}
\titlerunning{\NHmath-\NHI correlation in GPS galaxies}
\authorrunning{L. Ostorero et al.}
\institute{
       Dipartimento di Fisica, 
       Universit\`a degli Studi di Torino, Via P. Giuria 1, 
       10125 Torino, Italy
\and 
       Istituto Nazionale di Fisica Nucleare (INFN), Via P. 
       Giuria 1, 10125 Torino, Italy
\and  
      Netherlands Institute for Radio Astronomy, Postbus 2, 7990 AA Dwingeloo, 
      The Netherlands
\and  
      Kapteyn Astronomical Institute, University of Groningen, P.O. Box 800, 
      9700 AV Groningen, The Netherlands
\and  
       Harvard-Smithsonian Center for Astrophysics, 60 Garden St., Cambridge, MA 02138, USA
\and
       Astronomical Observatory, Jagiellonian University, ul. Orla
       171, 30-244  Krak\'ow, Poland
\and
       Nicolaus Copernicus Astronomical Center, Bartycka 18, 00-716
       Warsaw, Poland
\and
      Institute for Astronomy, Department of Physics, ETH Zurich, CH-8093 Zurich, Switzerland
}

\received{XXXX}
\accepted{XXXX}
\publonline{XXXX}


\abstract{With the Westerbork Synthesis Radio Telescope, we performed \HI observations 
of a sample of known X-ray emitting Gigahertz-peaked-spectrum galaxies with compact-symmetric-object 
morphology (GPS/CSOs) that lacked an HI absorption detection.
We combined radio and X-ray data of the full sample of X-ray emitting GPS/CSOs and found a significant, 
positive correlation between the column densities of the total and neutral hydrogen (\NH and \NHImath, 
respectively). 
Using a Bayesian approach, we simultaneously quantified the parameters of the 
\NH - \NHI relation and the intrinsic spread of the data set.
For a specific subset of our sample, we found \NHmath$\propto$\NHImath$^{b}$, with $b=0.93^{+0.49}_{-0.33}$, 
and $\sigma_{\rm int}($\NHmath$)= 1.27^{+1.30}_{-0.40}$. 
The \NH - \NHI correlation suggests a connection between the physical properties of the radio and X-ray absorbing gas.}

\maketitle

\section{Introduction}

Compact radio galaxies with GHz-peaked spectrum (GPS) and compact-symmetric-object 
(CSO) morphology, hereafter referred to as GPS/CSOs, are powerful, sub-kiloparsec scale 
sources that likely represent the youngest fraction of the radio galaxy population (e.g., O'Dea 1998; 
Orienti 2015).
Their radio emission is dominated by the mini-lobes, fully contained in the host galaxy interstellar medium.
Although they are increasingly detected in the X-ray domain (O'Dea et al. 2000; Risaliti et al. 2003; 
Guainazzi et al. 2004, 2006; Vink et al. 2006; Siemiginowska et al. 2008; Tengstrand et al. 2009), 
the best angular resolution currently available ($\sim$1$''$ with {\it Chandra}) is not sufficient to 
resolve the X-ray morphology of most GPS/CSOs. Therefore, the origin of their X-ray emission 
is still matter of debate (e.g., Siemiginowska et al. 2009; Migliori 2015). 

Both the radio and the X-ray emission display significant absorption. However, the location of 
the radio and X-ray absorbers in the host galaxy of the radio source are still debated too.
In the limited sample of compact sources with available high angular resolution \HI absorption measurements, 
the \HI is typically detected against one or both radio lobes (Araya et al. 2010, and references therein; Morganti et al. 2013),
with estimated covering factors, $C_{\rm f}$, variable from $\sim$0.2 to $\sim$1 (e.g., Peck et al. 1999; Morganti et al. 2004). 
The general consensus is that the \HI absorber has an inhomogeneous or clumpy nature, although the actual distribution 
of the gas is not known.

Hard X-ray (2--10 keV) spectra of GPS/CSOs reveal a  mean value of the column density 
much higher than that of a control sample of exended radio galaxies of the FR-I type
(whose core emission appears to be generally unobscured by an optically thick torus), 
and intermediate between the values of unobscured and highly obscured 
FR-II radio galaxies, where the presence of an optically thick torus is supported by 
optical and X-ray observations (Tengstrand et al. 2009, and references therein).
Any relationship between the radio and X-ray absorbers may thus help to clarify the absorption picture.

A comparison between \NH and \NHI for a sample of GPS/CSOs showed that the \NH values 
are systematically higher than the \NHI values of 1--2 orders of magnitudes (Vink et al. 2006; Tengstrand et al. 2009).
Because the estimate of \NHI \\ depends on the ratio between the spin temperature of the absorbing gas and its  
source covering factor, $T_{\rm s}/C_{\rm f}$ (see Equation 2), the magnitude of this offset can change according
to the  $T_{\rm s}/C_{\rm f}$ parameter, which is poorly constrained (Ostorero et al. 2010, and references therein; 
Curran et al. 2013).
Regardless of its magnitude, however, the existence of a systematic offset 
might in itself indicate a connection between X-ray and radio absorbers, and 
deserves to be investigated statistically.

In a sample of 10 GPS/CSOs, whose \NH and \NHI measu\-rements were available in the literature,
we discovered a significant, positive \NH-\NHI correlation (Ostorero et al. 2009, 2010),
with \NH and \NHI linked through the relationhsip \NH $\propto$ \NHI$^b$, with $b\simeq 1$.
With the aim of improving the statistics of the \NH-\NHI correlation sample,  
we carried out a program of observations of the known X-ray emitting GPS/CSOs lacking an \HI absorption detection
with the Westerbork Synthesis Radio Telescope (WSRT). In this paper, we report the first results of this project.

\section{Source sample}

We studied the \HI and X-ray absorption properties of the full sample of GPS galaxies 
known as X-ray emitters to date.
This sample is composed of 21 sources, belonging to at least one of the 
following GPS samples: Stanghellini et al. (1998), Snellen et al. (1998), Torniainen et al. (2007), 
and Vermeulen et al. (2003). All of these sources were also classified as CSOs.
The source list is given in Table \ref{tab_sources}.

\begin{table}
\caption{GPS/CSO sources of our sample: source names, optical redshifts, type of \NHI data (from spatially unresolved 
measurements) and of \NH data included in the present analysis (V: value, UL: upper limit, NA: not available 
because of bad data quality, IP: analysis in progress) and corresponding references ($\star$:~this 
work, and Ostorero et al., in prep.; $\star\star$: Siemiginowska et al., in prep.; numbers refer to papers cited in 
the reference list, where they are included in square brackets).}
\label{tab_sources}
\begin{tabular}{llllll}\hline
Source name  & $z_{\rm opt}$ & \NHI & Ref.              & \NH   	& Ref.    \\ 
(B1950)     & ~~           & ~~   & ~~    		& ~~    	& ~~      \\
\hline
0026+346    & 0.517    	  & NA   & $\star$      	& V 		& 9  \\
0108+388    & 0.66847     & V  	 & 1    		& LL      	& 10  \\
0116+319    & 0.06     	  & V  	 & 2    		& IP 		& $\star\star$ \\
0428+205    & 0.219 	  & V  	 & 3      		& UL 		& 10  	\\
0500+019    & 0.58457	  & V  	 & 4      		& V 		& 10 \\
0710+439    & 0.518 	  & NA 	 & $\star$ 		& V 		& 10 \\
0941-080    & 0.228 	  & V  	 & $\star$	    	& UL 		& 11,12 \\
1031+567    & 0.450 	  & V 	 & $\star$	    	& V 		& 10 \\
1117+146    & 0.362 	  & UL 	 & $\star$		& UL  		& 10  	\\
1323+321    & 0.370 	  & V  	 & 3      		& V 		& 10  	\\
1345+125    & 0.12174	  & V  	 & 5 			& V 		& 10,12\\
1358+624    & 0.431 	  & V  	 & 3      		& V 		& 10  	\\
1404+286    & 0.07658	  & V  	 & 1     		& V 	        & 13 \\  
1607+268    & 0.473 	  & UL 	 & $\star$   		& UL	        & 10	\\
1843+356    & 0.764 	  & UL 	 & $\star$ 		& IP  		& $\star\star$ 	\\
1934-638    & 0.18129	  & V  	 & 6      		& UL		& 14 	\\
1946+708    & 0.101 	  & V  	 & 7,8     		& V      	& 14 	\\
2008-068    & 0.547 	  & NA 	 & $\star$	     	& UL  		& 10 	\\
2021+614    & 0.227 	  & UL 	 & $\star$		& IP  		& $\star\star$ 	\\
2128+048    & 0.99 	  & UL 	 & $\star$    		& V,UL		& 9,10 	\\ 
2352+495    & 0.2379 	  & V  	 & 3      		& V  		& 10	\\
\hline
\end{tabular}
\end{table}

\section{HI observations with the WSRT}

With the WSRT, we searched for \HI absorption the 10 sources of our sample (marked with a star in Column 4 of 
Table \ref{tab_sources}) that still lacked an \HI detection,
either because they were never observed or because previous observations 
yielded only upper limits to the HI optical depth.

Each target was observed for an exposure time of four to 12 hours with the UHF-high-band receiver (appropriate 
when $z \gtsim 0.2$) in dual orthogonal polarisation mode. 
The observing band was 10 to 20 MHz wide, with 1024 spectral channels, and was centered at the 
frequency where the \HI absorption line is expected to occur based on the optical redshift.
Compared to the \HI survey of compact sources by Vermeulen et al. (2003),
our observations could benefit from  a larger ratio between number of spectral channels and observing 
band width: this improvement, together with the longer exposure times, enabled a more effective 
separation of narrow \HI absorption features from radio frequency interferences (RFI) in most of 
the previously observed sources.
The data reduction process and the results of the analysis will be presented elsewhere.

We detected \HI absorption in two out of 10 targets. 
For five targets, we could estimate upper limits to the line optical depth and
to the \HI column density. For three of our targets, RFI were too strong to obtain any useful data 
(see Table \ref{tab_sources}).

The atomic hydrogen column density along the line of sight is related to the velocity integrated optical 
depth of the 21-cm absorption via the following relationship (Wolfe \& Burbidge 1975): 

\begin{equation}
\label{eq_NHI}
$\NHI$=1.823\times 10^{18}\, T_{\mathrm s} \int{\tau \,{\rm d}v} .
\end{equation}
In the optically thin regime (i.e. for $\tau \ltsim 0.3$), Equation \ref{eq_NHI} is approximated by
\begin{equation}
\label{eq_NHI_appr}
$\NHI$ \approx 1.823\times 10^{18}\, (T_{\mathrm s}/C_{\mathrm f}) \int{\tau_{\rm obs}\, {\rm d}v} ,
\end{equation}
where $\tau_{\rm obs} \equiv \Delta S/S$ is the observed optical depth of the line, given by 
the ratio between the spectral line depth ($\Delta S$) and the continuum flux ($S$) of the background radio 
source, and is related to the actual optical depth through $\tau\approx \tau_{\rm obs}/C_{\mathrm f}$.
For homogeneity with literature data, we estimated the \HI column densities by assuming that the absorber  
fully covers the radio source ($C_{\mathrm f}$=1) and that the spin temperature is $T_{\mathrm s}$=100 K.
These assumptions yielded \NHI values that are lower limits to the actual column densities. 
When no \HI line was detected, upper limits to \NHI were estimated from the optical depth 
upper limits ($\tau_{3\sigma}$) that we derived from the 3-$\sigma$ noise level.

\section{\NH - \NHI correlation}

We investigated the existence of a correlation between the neutral 
hydrogen column density (\NHImath) and the total hydrogen column density (\NHmath),
with the aim of either confirming or disproving the significant, positive \NH-\NHI  
correlation that we discovered for a smaller sample of 10 GPS/CSOs (Ostorero et al. 2009, 2010).
Below, we describe the selection of the correlation sample and present the first results of the correlation analysis.

\subsection{\NHI sample}

\HI absorption detections and corresponding \NHI values 
were available for 13 out of 21 sources of the sample; 
\NHI upper limits could be estimated for five sources; 
for three sources, no useful data were obtained 
(see Table \ref{tab_sources}).

Because multiple spectral features were detected in the \HI absorption spectra 
of a subsample of sources, different \NHI values could be associated to the same source.
Here, we present the result of the analysis that we performed by taking 
the {\it total} \NHI of the detected absorption features into account. 
The full record of cases will be presented elsewhere.
When more than one total \NHI estimate was available for a given source, we chose the most recent result 
from spatially unresolved measurements.

\subsection{\NH sample}
\label{sec_nh_sample}

Estimates of the total hydrogen column densities (\NHmath) were available for 18 out of 21 sources:
11 of them are \NH values,
six of them are upper limits, and one of them is a lower limit to \NH. 
For the remaining sources, the analysis is in progress 
(see Table \ref{tab_sources}).

The \NH estimates were derived from the X-ray spectral analysis by fitting a model spectrum, 
absorbed by both a Galactic and a local (i.e., at redshift $z_{\mathrm opt}$) gas column density, 
to the observed X-ray spectrum; in the fitting procedure, the \NH parameter was left free to vary.
For some sources, more than one model could be fit to the spectrum, yielding different \NH best-fit estimates. 
In particular, four sources (1404+286, 1607+268, 1934-638, and 1946+708) could satisfactorily 
be interpreted as both Compton-thin and Compton-thick X-ray sources:
here, we present the analysis performed by taking the Compton-thin \NH values 
only into account; the full record of cases will be discussed elsewhere.
Furthermore, for one source (0108+388), the available lower limit to \NH is consistent with both a Compton-thin 
and a Compton-thick scenario: again, here we present the results of the analysis performed under the assumption that 
the source is Compton-thin. 
For five sources of the sample,  
different \NH estimates were derived from observations carried out in different epochs.
Because we cannot rule out long-term column-density variations, 
when different values 
were not consistent with each other at the 1-$\sigma$ level (as for 1345+125)
and when both values 
and upper limits were available (as for 2128+048),
we chose to include the two extreme \NH estimates in our correlation sample, 
each of them associated to the same \NHI value.
Including either the average or any of the two \NH estimates does not change our results significantly.

\subsection{\NH - \NHI sample}

Pairs of \NHI and \NH {\it values} 
(V-V pairs in Table \ref{tab_sources}) were available for 
a subsample of eight sources, hereafter referred to as the {\it detection correlation sample}, $D$.
We also define a second {\it detection correlation sample},  $D'$, which includes, besides the sources 
of sample {\it D}, source 0108+388,
for which a proper detection is not available. Under the assumption that the source is Compton thin 
(see Section \ref{sec_nh_sample}) 
an \NH value can be estimated as the mean of the 3-$\sigma$ lower limit 
and the physical upper bound of the Compton-thin \NH range (i.e., \NH=$9.0\times 10^{23}$ \percmq). 
We consider the \NH range as a $\pm 3\sigma$ interval, and thus associate to the mean \NH the corresponding 
1-$\sigma$ error.
Pairs of \NHI and \NH {\it estimates} 
(i.e., pairs including detections and/or upper limits: V-V, V-UL, UL-V, and UL-UL pairs in Table 
\ref{tab_sources}) were available for a subsample of 14 sources, hereafter referred to as the 
{\it estimate correlation sample}, $E$.
As we did for the detections, we also define a second {\it estimate correlation sample}, $E'$, which includes sample 
$E$ and source 0108+388.

\subsection{Correlation analysis}
\label{subsec_correlation}

\begin{figure}
{\includegraphics[scale=0.49]{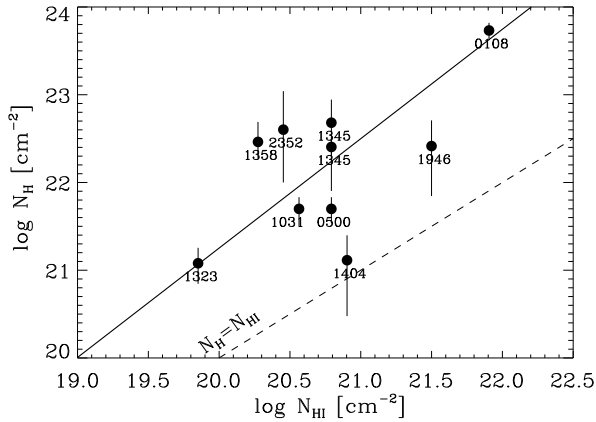}}
\caption {X-ray column densities (\NHmath) as a function of radio column densities (\NHImath)
for the $D'_1$ sample. 
The solid symbols show the (\NHImath,\NHmath) measurements with  1-$\sigma$ error bars on \NH. 
\NHI was computed by assuming $T_{\mathrm{s}}=100$ K and $C_{\mathrm f}=1$.
Labels show a shortened version of the source names reported in Table \ref{tab_sources}.
Solid line: linear regression best-fit line to the data (to guide the eye; 
$\log $\NHmath$=a+b \log $\NHImath, with $a=-3.68$
and $b=1.25$);
dashed line: bisector of the \NHImath-\NH plane. Data are from references listed in Table \ref{tab_sources}.}
\label{fig1}
\end{figure}

\begin{figure}
{\includegraphics[scale=0.49]{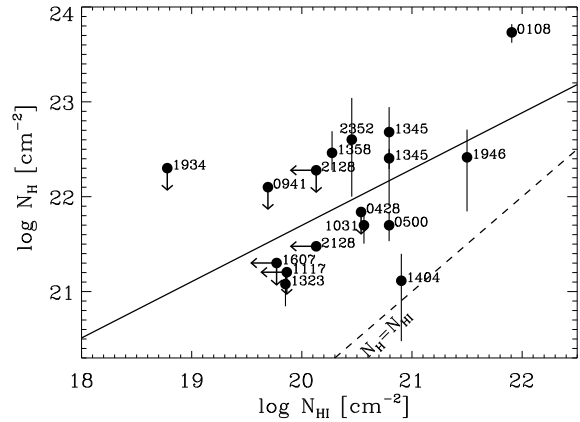}}
\caption{X-ray column densities (\NHmath) as a function of radio column densities (\NHImath)
for the $E'_1$ sample.
The solid symbols show the (\NHImath,\NHmath) measurements with  1-$\sigma$ error bars on \NH. 
\NHI was computed by assuming $T_{\mathrm{s}}=100$ K and $C_{\mathrm f}=1$.
Arrows represents upper limits. Labels show a shortened version of the source names reported in Table \ref{tab_sources}.
Solid line: linear regression best-fit line to the data (to guide the eye; 
$\log $\NHmath$=a+b \log $\NHImath, with $a=-9.8$ and $b=0.59$);
dashed line: bisector of the \NHImath-\NH plane. 
Data are from references listed in Table \ref{tab_sources}.}
\label{fig2}
\end{figure}

Figures \ref{fig1}  and \ref{fig2}  display 
the (\NHImath,\NHmath) data for the {\it detection} and {\it estimate correlation samples}, respectively.

First, we performed the correlation analysis on the  {\it detection correlation sample}.
A Pearson correlation analysis applied to sample $D$ did not 
reveal a significant correlation. 
This result is confirmed by the more robust, {\it non-parametric} (or {\it rank}) 
methods, i.e.\ Spearman's and Kendall's correlation analysis (Press 1992). 
The main reason why we do not confirm the result found for the detections in our previous paper
(Ostorero et al. 2009, 2010) is that we replaced the \NH values that were derived by fixing 
the X-ray spectral index (Vink et al. 2006)  
with more robust estimates (Tengstrand et al. 2009). 
We also chose to include source 1404+286 (with a complex X-ray spectrum) in our sample.

We then repeated the Pearson correlation analysis for sample $D'$, and we find a positive, significant
correlation (correlation coefficient: $r=0.93$; probability of the null-hypothesis of no correlation being true: $P = 7.8\times 10^{-5}$).
The statistical significance of this correlation, however, is admittedly 
driven by source 0108+388, characterized by the largest (\NHImath,\NHmath) values. 
Indeed, in the same sample, the evaluation of the correlation by means of the more robust Spearman's 
and Kendall's correlation analysis dramatically decreased the significance of the above result ($P=0.3$).

In the {\it estimate correlation sample}, we investigated the correlation by means of survival analysis techniques. 
In particular, we made use of the software package ASURV Rev.\ $1.3$\footnote{\tt http://www2.astro.psu.edu/statcodes/asurv/} (Lavalley, 
Isobe \& Feigelson 1992), which implements the methods for bivariate problems presented in Isobe, Feigelson \& Nelson (1986).
The ASURV generalized Spearman's and Kendall's correlation analysis applied to sample {\it E} shows that the data are significantly 
correlated ($P= 1.8\times 10^{-2} - 3.8\times 10^{-2}$). 
For sample {\it E'}, including source 
0108+388\footnote{for 0108+388, we still use the \NH value estimated as described in Section \ref{sec_nh_sample}, 
instead of the lower limit, because ASURV cannot deal with two types of data censoring.}
the significance improved further ($P= 6.5\times 10^{-3} - 9\times 10^{-2}$).

As far as the law describing the relationship between \NH and \NHI is concerned, 
according to Pearson's test we could fit a linear relation to the {\it detection sample} $D'$:
 $\log$\NHmath$=a+b \log$\NHImath, 
with 
$a=-3.68\pm1.72$ and $b=1.25\pm0.08$.
However, this relation is not a good description of the data ($\chi^2_{\mathrm{red}}=6.73$):
the dispersion of the data is clearly larger than the typical uncertainties on \NH  
(see Figure \ref{fig1}).

As for the {\it estimate samples}, $E$ and $E'$, the ASURV Schmitt's linear regression enabled us to 
perform a linear fit to the data (the estimated best-fit slope is $b=0.59$) 
but not to evaluate the goodness of the fit.
A visual inspection of this data set again confirms a dispersion larger than the typical \NH uncertainties (see Figure \ref{fig2}).
This fact suggests that additional variables are involved in the correlation, and the \NH - \NHI relation 
is the two-dimensional projection of a more complex relation.

\subsection{Bayesian analysis}

\begin{figure}
{\includegraphics[scale=0.49]{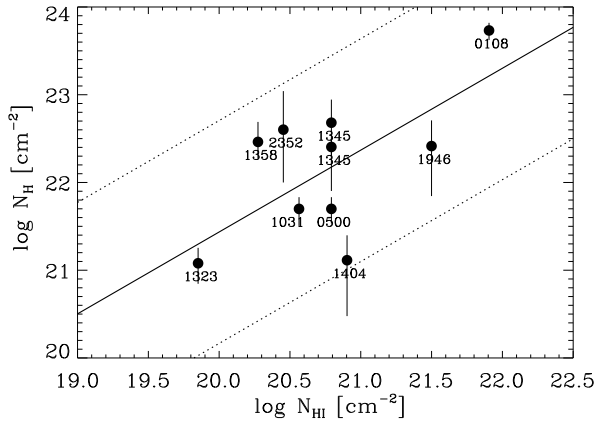}}
\caption{Bayesian analysis applied to sample $D'_1$. 
The solid symbols show the (\NHImath,\NHmath) measurements with  1-$\sigma$ error bars on \NH. 
The solid, straight-line is the \NH - \NHI relation: 
$\log$\NHmath$=a+b\log$\NHImath, with $a=2.78$ and $b=0.93$.
The dotted lines show the $\pm\sigma_{\rm int}$ standard deviation of the relation, with 
$\sigma_{\rm int}= 1.27$.
\NHI was computed by assuming $T_{\mathrm{s}}=100$~K and $C_{\mathrm f}=1$.}
\label{fig3}
\end{figure}

In order to derive the correlation parameters of the \NH - \NHI relation and the intrinsic scatter of the data set, 
it is appropriate to resort to Bayesian analysis.
Given our data set
$DS=\{x^i,y^i,S_1^i,S_2^i\}$, 
where $\{x^{i}\}=\{\log$\NHImath$^i\}$ and $\{y^i\}=\{\log$\NHmath$^i\}$ are 
the vectors of the $\log$\NHI and  
$\log$\NH \\
values, respectively, 
$\{S_1^i\}$ and $\{S_2^i\}$ are the vectors of the upper and lower uncertainties on the $\{y^i\}$ values,
and $i$ runs from 1 to the number of data points, $N$,  
we determined the multi-dimensional probability density function (PDF) of the parameters 
$\theta=\{a,b,\sigma_{\rm int}\}$. Here, $a$ and $b$ are the parameters of the correlation: 
$y = a + bx$. 
In order to mimic additional hidden parameters in the relation, we assumed 
that $y^i$ is a random variate with mean
$\bar y^i= a + b x^i $
and variance $\sigma_{\rm int}^2$,  
with $\sigma_{\rm int}$ the intrinsic scatter of the 
dependent variable (e.g., Andreon \& Hurn 2010).

For our Bayesian analysis, we used the code APEMoST\footnote{Automated Parameter Estimation and Model
Selection Toolkit; {\tt http://apemost.sourceforge.net/}, Feb. 2011.}, developed by 
J. Buchner and M. Gruberbauer (e.g., Gruberbauer 2009).
Because this code cannot deal with censored data, we restrict our Bayesian analysis to the {\it detection sample} $D'$. 
We assumed independent flat priors for parameters $a$ and $b$, except for the internal 
dispersion $\sigma_{\rm int}$, which is positive defined.
We use $2\times10^6$ MCMC iterations to guarantee a fairly complete sampling of the parameter
space.
The boundaries of the parameter space
were set to $[-1000, 1000]$ for the $a$ and $b$ parameters, $[0.01, 1000]$ for the $\sigma_{\rm int}$ parameters. 
The initial seed of the random number generator was set with the 
{\tt bash} command {\tt \verb"GSL_RANDOM_"}\\ {\tt \verb"SEED=$RANDOM"}.

Our analysis shows that, for any given value of $\log$\NHImath, $\log$\NH takes the value $\log$\NH$=a+b\log$\NHI$\pm\sigma_{int}$ 
(with $a=2.78^{+6.82}_{-9.83}$, $b=0.93^{+0.49}_{-0.33}$, and $\sigma_{\rm int}=1.27^{+1.30}_{-0.40}$) 
with a 68 percent probability. 
As an example, when \NHI is equal to $10^{20}$ \percmq, \NH varies in the interval ($10^{20.11}-10^{22.65}$) \percmq.
Although both $\sigma_{\rm int}$ and the uncertainties on 
$a$, $b$, and $\sigma_{\rm int}$  
are large, the \NH - \NHI relation derived from the Bayesian analysis properly describes our data set, 
unlike the linear relations derived in Section \ref{subsec_correlation}.
The results of the Bayesian analysis are displayed in Figure \ref{fig3}.

\section{Conclusions}
We performed spatially unresolved \HI absorption observation of a sample of GPS/CSO galaxies
with the WSRT, with the aim of improving the statistics of the \NH-\NHI correlation sample.
We confirmed a significant, positive correlation between \NH and \NHI. 
For the full, censored data set, the generalized linear regression analysis yields 
\NHmath$\propto$\NHImath$^b$, with $b\simeq 0.6$, although the goodness of this fit cannot be evaluated.
For the detection subset, the linear regression analysis yields 
\NHmath$\propto$\NHImath$^b$, with $b\simeq 1$, however the $\chi^2$ is large.
This fact may indicate that the \NH - \NHI relation is not a one-to-one relation: additional variables are involved in 
the correlation and generate the intrinsic spread of the data set.
The Bayesian analysis applied to the detection subset enabled us to simultaneously quantify (i) the parameters 
of the \NH - \NHI relation, whose slope is $b=0.93^{+0.49}_{-0.33}$, and (ii) the intrinsic spread on \NH, 
$\sigma_{\rm int}=1.27^{+1.30}_{-0.40}$. 
The \NH - \NHI correlation suggests a connection between the intrinsic properties of the X-ray and \HI absorbers.
We are currently investigating the implications of this correlation, and we are attempting to 
identify and constrain the additional variables responsible for the intrinsic spread.

\acknowledgements
We thank the organisers of the Fifth Workshop on CSS and GPS sources for an extremely 
stimulating and enjoyable meeting.
The research leading to these results has received funding from the European Commission 
Seventh Framework Programme (FP/2007-2013) under grant agreement No 283393 (RadioNet3).
L.O. acknowledges support from the ``Helena Kluyver'' programme 
run by ASTRON/JIVE, the INFN grant INDARK, the grant PRIN 2012 ``Fisica Astroparticellare Teorica''
of the Italian Ministry of University and Research, 
and the ``Strategic Research Grant: Origin and Detection of Galactic and Extragalactic 
Cosmic Rays'' funded by the University of Torino and Compagnia di San Paolo.
L.O. is grateful to the Department of Physics and Astronomy of the University of Pennsylvania
and to the High Energy Astrophysics Division of the Harvard-Smithsonian Center for Astrophysics
for their support and kind hospitality. 
\L.S. was supported by the Polish National Science Centre through the grant DEC- 2012/04/A/ST9/00083.
We are grateful to G.J\'ozsa for his support during the observing runs with the WSRT and for 
providing us with the data-cubes of our target sources.
We thank J.Buchner and M.Gruberbauer for developing their superb
code APEMoST and making it available to the community.
S.Andreon is acknoweldged for a very stimulating seminar on Bayesian statistics.
The WSRT is operated by ASTRON (Netherlands Institute for Radio Astronomy) with support from the 
Netherlands Foundation for Scientific Research (NWO).

\end{document}